
\magnification=\magstep1
\def\refs{\leftskip=.3truein\parindent=-.3truein}
\def\endrefs{\leftskip=-.3truein\parindent=.3truein}
\baselineskip=24pt

\centerline{ENERGY PRINCIPLES FOR SELF-GRAVITATING BAROTROPIC FLOWS:}
\centerline{I. GENERAL THEORY}

\bigskip

\centerline{Joseph Katz\footnote*{Permanent address:~~ The Racah Institute of
Physics, Jerusalem 91904, Israel.}}

\centerline{National Astronomical Observatory, Mitaka, Tokyo 181}

\centerline{Shogo Inagaki}

\centerline{Department of Astronomy, Faculty of Science, Kyoto University,}

\centerline{Sakyo-ku, Kyoto 606-01}

\centerline{and}

\centerline{Asher Yahalom}

\centerline{The Racah Institue of Physics, Jerusalem 91904, Israel}
\vfill\eject

\centerline{\bf Abstract}
\noindent
The following principle of minimum energy may be a powerful substitute to the
dynamical perturbation method, when the latter is hard to apply.
Fluid elements of self-gravitating barotropic flows, whose vortex lines
extend to the boundary of the fluid, are labelled in such a way
that any change of trial configurations automatically preserves mass and
circulation.
The velocity field is given by a mass conserving
Clebsch representation. With three independent Lagrangian functions, the total
energy is
stationary for all small variations about a flow with  fixed linear and angular
momenta provided
Euler's equations for steady motion are satisfied.  Thus, steady flows
are stable if their energy is minimum.  Since energy is here  minimized
subject to having local and global contants of the motion fixed,
stability limits obtained that way are expected to
be close to limits given by dynamical perturbation
methods. Moreover, the stability limits are with respect to arbitrary, not
necessary small, perturbations.
A weaker form of the energy principle is also given which may be easier to
apply.

The Lagrangian functional, with the same three Lagrange variables is stationary
for
the fully time dependent Euler equations. It follows that the principle of
minimum
energy gives stability conditions that are both necessary and sufficient
if terms linear in time derivatives (gyroscopic terms)
are absent from the Lagrangian.  The gyroscopic term for small deviations
around steady  flows is given explicitly.

\vfill\eject

\noindent
{\bf 1. Introduction}
\bigskip

 The method of small perturbations and analyses of eigenmodes are
 difficult to apply to astrophysical models that are not at least
 axisymmetric (Binney \& Tremaine 1987). Moreover, it gives  good
 indications on how instabilities set in, but once a
 configuration has become unstable, the method is at a loss to say
 whether it tends towards a new steady figure or
 will oscillate like a pendulum or become chaotic.
 In this era of supercomputers, one might think that a dynamical
 follow-up is  routine work, but
 this is not the case, yet. Moreover, as a rule,
analytical methods give more
 insight than purely numerical ones.
 Therefore, there
 are still reasons for perfecting other methods like the principle of
 minimum energy which played important roles in stability theory [see
in particular  Antonov (1962), Lebovitz (1965) and  Chandrasekhar's (1969)
works].

 We shall be concerned with ideal fluids and, for definiteness,
 we limit ourselves to barotropic fluids. In our Lagrange variables,
 non-barotropic fluids are more or less a particular case than a different one
 and may be treated along the same lines. We shall not do this here.

 From the pioneering work of Arnold (1966), see also Lynden-Bell \& Katz
 (1981) on compressible flows (the paper will be referred to below as
 LBK81) we know that among all steady flows that are mass preserving
 and isocirculational, those whose energy is stationary for small variations
satisfy Euler's
equations. By mass preserving we mean that  in an arbitrary virtual
displacement of a small element of fluid there is no loss of mass
and by isocirculational we mean that
 the flux of the vortex lines, in any direction through a small area, is the
 same for the displaced element of fluid.

 It follows that if a steady configuration has minimum energy, it
 is stable with respect to perturbations that are not necessarily
 small.  In a perturbation analysis, perturbations are always assumed to be
 small.  Thus,
 while there may be stable configurations with respect to small perturbations
 for which the energy is not a minimum, the stability limits given by the
 energy principle are stronger than those of a perturbation analysis.

This principle of minimum energy can be made even stronger by imposing further,
as
constraints, global invariants of the motions.  In
that way, virtual displacements may eventually be restricted as much as real
motions.  If, by varying some parameter, a series of stable configurations
ceases to have minimum energy with such severe constraints, the limit may
be quite close to the limit one would find in a perturbation analysis.

Another way of using the \lq\lq maximally constrained" energy consists in using
a computer to find a new minimum when the energy principle fails to indicate
stability. Since time is not
involved,  calculations have one dimension less than the dynamical equations ;
this may be of great computational help. The new configuration, if any,
will be one in which the unstable model is likely to settle.
 The absence of a new steady configuration may indicate that no
 new real stable configuration exists.

A key point in applying this energy principle is to make an appropriate
parametrization of the fluid elements; this includes the topology of the flows
that must be conserved as well.  The parametrization used here has been found
in  LBK81 where labels are attached to vortex lines rather than to fluid
elements.  Such a labelling leads naturally to a particular Clebsch
representation for the velocity of the flow.  A Lagrangian and Energy
Principle in terms of three independent  functions was developped in Katz and
Lynden-Bell(1985)  (For a formal generalization of that approach, see
Simo,Lewis and Marsden 1991).  The method contains no Lagrange multiplier as is
usually
done [Serrin 1959, Lin 1963, Seliger and Whitham 1968] as such multipliers are
not very usuful in second variations.

The new elements in this work, as compared to LBK 81 and KLB 85, are as
follows.
We specify completely the labelling in a class of flows.  We use a new set of
Lagrangian functions to prove
energy principles for steady motions.  We show the uniqueness of our choice
and fix whatever freedom remains in positioning the coordinates.  The principle
of minimum energy is completed with the proof that our variables are true
Lagrangian variables. That is, Euler's dynamical equations are obtained by
varying the Action.  If the Lagrangian's kinetic energy is purely
quadratic in time derivatives (no gyroscopic term),
the principle of energy minimum is a necessary
and sufficient condition for stability.  The form of the terms linear in time
derivatives gives thus important information and have therefore been worked out
explicitly.  The formulation follows closely
classical mechanics.  Applications to two dimensional flows and in particular
to MacLaurin disks, whose dynamical stability limits are well known, give a
perfect illustration of the power of our method.  To maintain this paper
within reasonable limits, we delay the application to an accompanying paper II.
\bigskip
\noindent
{\bf 2. Stationary Barotropic Self-Gravitating Flows}
\bigskip

 	It will be useful to write the equations of motion
 first to show our essentially standard notations:
${\vec r} =(x^k)= (x,y,z) ~~[k,l,m,n = 1,2,3]$ for the
 positions, ${\vec W}$ for the steady velocity field in
 inertial coordinates, $\rho$  the  density, $P(\rho)$
 the pressure, $h(\rho) = \int dP/\rho$ the specific enthalpy and $\Phi (<0)$
 the gravitational potential,
 $$\Phi= - G \int {\rho(\vec r^{~'})\over{|\vec r - \vec r^{~'}| }}d^3 x'.
    \eqno(2.1)$$
Euler's equations for flows that are steady in coordinates with uniform
velocity ${\vec b}$  and angular velocity\footnote*{The index c of  $\vec
\Omega_c$ referes to rotating coordinates and is just here to distinguish it
from $\vec \Omega$ with no index, commonly used for angular velocities in
galactic disks or rotating ste}
velocity ${\vec U}$ may be written
$$ \vec {\cal O} \equiv (\vec U \cdot \vec \nabla)\vec W +
   \vec  \Omega_c \times \vec W + \vec \nabla (h + \Phi) = 0,  \eqno(2.2)$$
where, by definition,
$$ {\vec W} ={\vec U} + {\vec b} + {\vec  \Omega_c} \times {\vec r} \equiv \vec
U + \vec \eta_c.
    \eqno(2.3)$$
Another useful form of Euler's equation is
$$\vec {\cal O} = \vec \omega \times \vec U + \vec \nabla \Lambda = 0,
   \eqno(2.4)$$
where
$$ \vec \omega = {\rm rot}~ \vec W         \eqno(2.5)$$
and
$$ \Lambda = {1 \over 2} (\vec U^{2} - \vec \eta_c^2) + h + \Phi.
\eqno(2.6)$$
With (2.4) we also have the equation of mass conservation
$$ {\cal U} \equiv \vec \nabla\cdot (\rho \vec U) = 0.      \eqno(2.7)$$
The equation of circulation conservation follows from (2.4):
$$ {\rm rot}~(\vec \omega \times \vec U) = 0.         \eqno(2.8)$$
The boundary conditions are those of a self-gravitating flow in free space,
the density and the pressure go to zero and
the velocity of sound goes to zero as well:
$$\rho |_s  = 0 \qquad P |_s = 0  \qquad {dP \over d \rho}\bigg|_{\rho=0}
   =0.   \eqno(2.9)$$
Notice that (2.7) with (2.9) implies
$$ \vec U \cdot \vec \nabla \rho |_s = 0, \eqno(2.10)  $$
i.e., $\vec U$ is in the tangent plane of the surface of the fluid.
Apart from satisfying these boundary conditions, physical quantities are
assumed  to be bounded everywhere.

In {\it time dependent flows}, in addition to mass conservation and
conservation of
circulation, the total mass, the linear and the angular momentum are conserved;
the center of mass moves with a uniform velocity or is at rest.
The conserved global quantities are
$$ M = \int \rho d^3x, \qquad \vec P = \int \vec W \rho d^3 x,  \qquad
   \vec J = \int \vec r \times \vec W \rho d^3x.             \eqno(2.11)$$
It is always possible and worthwhile to take
$$ {\vec P} = 0. \eqno(2.12)$$
What this has to do with steady flows will soon become clear.
\vfill\eject
\noindent
{\bf 3. Mass Preserving and Isocirculational Labelling}
\bigskip

\noindent
{\it 3.1. Labelling of Fluid Elements When the Vortex Lines Extend to the
Boundary}

Conservation of  vorticity implies conservation of the topology of the vortex
lines. Flows with given vorticity have thus also given topologies of
$\vec \omega$-lines. For definiteness  we shall consider flows with the same
topology as in LBK81 (see figure 1).  There, vortex lines were labelled by
three parameters: the load $\lambda$, the metage $\mu$ and an \lq\lq angle"
$\beta$.

Surfaces of constant load $\lambda$ are surfaces of constant mass per unit
vortex strength in a narrow tube of vortex lines: if $dM$ is the mass in a tube
with vorticity flux $dC$,
$$ \lambda = {d M \over d C} = \int_{\rm Bottom}^{\rm Top} {\rho \over \omega}
   d l.  \eqno(3.1) $$
The integration is along ${\vec \omega}$-lines. The surfaces  $\lambda  =
{\rm constant}$ are embedded \lq\lq cylinders''
as shown in figure 1. The cylinders may be parametrized in any way we want, say
$\alpha = \alpha ( \lambda) = const. $, but
the particular parametrization in which $\alpha (\lambda )$ is proportional to
the circulation on a closed contour around $\lambda =const. $  has very special
property (see below).

The metage $\mu$, is defined as another family of surfaces with the same
integral
as in (3.1) taken up to some \lq\lq red mark" on the ${\vec \omega}$-line:
$$  \mu = \int_{\rm Bottom}^{\rm Red~mark} {\rho \over \omega} dl.
    \eqno(3.2) $$
Surfaces of constant $\mu$ cut accross $\lambda$-surfaces. If the red mark is
on the bottom, $\mu=0$. If it is at the top,  $\mu = \lambda$.

The angular variable  $\beta$ is defined by a family of surfaces of
$\vec \omega$-lines as well ; these are  `\lq cuts'' hanging on a \lq\lq
central line" (see figure 1).  Thus, by definition,
$$ {\vec \omega} \cdot {\vec \nabla} \alpha = 0, \qquad
   \vec \omega \cdot  \vec \nabla \beta = 0 \qquad {\rm and} \qquad
\vec \omega \cdot \vec \nabla \mu = \rho.     \eqno(3.3) $$
The parametrization of $\beta$ may be so chosen that
 $$ {\vec \omega} = {\vec \nabla} \alpha \times {\vec \nabla} \beta.
    \eqno(3.4)$$
This defines $\beta$ for given $\vec \omega$ and $\alpha$ up to an additive
single valued function of $\alpha$, $B(\alpha)$.  This $B(\alpha)$ is
associated with the
freedom to take any cut as $\beta=0$.  Having chosen $\alpha, \beta, \mu$,
It also follows from (3.3) that we must have
$$ \rho =
   \vec \nabla \alpha \times \vec \nabla \beta \cdot \vec \nabla \mu
   \eqno(3.5)$$
or that
 $$ \rho =   {\partial (\alpha, \beta, \mu) \over \partial (x, y, z)}.
\eqno(3.6)$$
and thus
$$ {\partial (\alpha, \beta, \mu) \over \partial (x, y, z)}
   \ne 0   \eqno(3.7)$$
everywhere except on the surface of the fluid since $\rho|_s = 0$,
$$ \rho|_s = \vec \omega \cdot \vec \nabla \mu|_s = 0.  \eqno(3.8)  $$
With (3.4) we see that the velocity field has now a Clebsch form
$$ \vec W = \alpha \vec \nabla \beta + \vec \nabla \nu.         \eqno(3.9)$$
The function $\nu$ we define by the condition that mass be preserved i.e.
by equation (2.7).  The function $\nu$ is thus defined by an elliptic equation
since (2.7) can be written
$$ \vec \nabla \cdot (\rho \vec \nabla \nu) =
    \vec \nabla \cdot [\rho ( - \alpha \vec \nabla \beta
   + \vec \eta_c)].    \eqno(3.10)$$
Since $\rho|_s = 0$, a regular single valued
solution  $\nu(\vec r)$ (up to a constant)
must be unique when it exists.

The main tool in practical calculations will thereby be the Green function
$G(\vec r , \vec r^{~'})$, solution of
$$\vec \nabla\cdot (\rho \vec \nabla G) = \delta^3(\vec r - \vec r^{~'}).
   \eqno(3.11)$$
\bigskip

\noindent
{\it 3.2. Uniqueness of the Labelling}
\bigskip

Consider now the effect of a reparametrization of the fluid elements in a flow
with given $\rho(\vec r),~ \vec W(\vec r) $  and thus
$\vec \omega (\vec r)$ and given $\lambda(\alpha)$.  Let
$\tilde \alpha, \tilde \beta, \tilde \mu$ be another labelling.
Surfaces of constant load may be reparametrized in any way:
$\alpha \to \tilde \alpha(\alpha)$.  Since
$$\vec \omega = \vec \nabla \alpha \times \vec \nabla \beta
  = \vec \nabla \tilde \alpha \times \vec \nabla \tilde \beta
  = {d \tilde \alpha \over d\alpha}
  \vec \nabla \alpha \times \vec \nabla \tilde \beta
  \eqno(3.12)$$
We must change $\beta$ to
$$ \tilde \beta = {d \alpha \over d\tilde \alpha} \beta + B(\alpha ).
   \eqno(3.13)$$
With both $\mu = \tilde \mu = 0 $ on the bottom of the fluid, the
parametrization of $\mu$ for given $\rho$ and $\vec \omega$ is uniquely defined
 $$ \tilde \mu = \mu.       \eqno(3.14)$$
The corresponding change in $\nu $ is accordingly defined by
$$ \vec W = \alpha \vec \nabla \beta + \vec \nabla \nu
   = \tilde \alpha \vec \nabla \tilde \beta + \vec \nabla \tilde \nu
   \eqno(3.15)$$
or
$$ \vec \nabla (\tilde \nu - \nu) = \alpha \vec \nabla \beta -
   \tilde \alpha \vec \nabla \tilde \beta.  \eqno(3.16) $$
{}From (3.13) it follows that
$$ \tilde \nu = \nu - (\tilde \alpha {d\alpha \over d\tilde \alpha} - \alpha )
    \beta - \int \tilde \alpha dB(\alpha) + {\rm Const}.
   \eqno(3.17)$$
$\alpha$ is single valued but $\beta$ is not.  If, however, we define and
keep $\nu$
single valued, any reparametrization that preserves the single valuedness of
$\nu$ must satisfy the condition
$$ {\tilde \alpha} {d\alpha \over d \tilde \alpha} - \alpha = 0 \quad {\rm or}
   \quad {\tilde \alpha} = l\alpha,  \eqno(3.18)$$
where $l$ is a constant. The parametrization for which $\nu$ is single valued
is easily found. On the one hand $C(\lambda)$ is given by the flux of $\vec
\omega$ through a surface with boundary $\alpha$ = constant.
$$  C(\lambda) = \int \int {\vec \omega} \cdot d{\vec S} = \int \int d\alpha
   d\beta = \int ^\alpha_{\alpha_c}[\beta]d\alpha,  \eqno(3.19)$$
where $[ \beta]$ is the value of the discontinuity of
$\beta$ and $\alpha_c$ is the value of
$\alpha$ on the central line.  On the other hand $C(\lambda)$ is also equal to
the circulation of $\vec W$ along a contour on a $\alpha$ = constant surface:
$$ C(\lambda) = \oint_{\alpha}{\vec W} \cdot d{\vec r}
   = \oint_\alpha \alpha d\beta = \alpha [\beta].   \eqno(3.20)$$
So, (3.19) and (3.20) give
$$ {dC \over d\alpha} = [\beta] = [\beta] + \alpha {d[\beta] \over d\alpha}
    \eqno(3.21)$$
which means that
$$ [\beta] = q \quad {\rm and} \quad \alpha = {1 \over q} C(\lambda),
\quad q=const. \eqno(3.22)$$
The parametrization of $\alpha$ by the circulation along $\lambda$-tubes is
unique except again, as in (3.18),
for the multiplication constant $q$; with $q$ = 2$\pi$, the domains
of the Lagrange variables $\alpha, \beta, \mu$ is thus
$$ 0 \le \alpha \le \alpha_{M}\equiv C_{MAX} \slash 2\pi,
   \quad B(\alpha) \le \beta \le B(\alpha)+2\pi,
   \quad 0\le \mu \le \lambda(\alpha),    \eqno(3.23)$$
where $\alpha$ = ${C(\lambda) \over 2 \pi}$ {\it must be a given function of
the flow}.
$C(\lambda)$ is similar to the amplitude  $J$ of the angular momentum. The only
remaining arbitrariness of the parametrization is $B(\alpha)$ associated with
the cut where $\beta = 0$.   We have thus found that the parametrization
$\alpha = C/2\pi$  of LBK81 is the only one (up to $q$) that
insures {\it single valued} $\nu$'s  in a Clebsch representation.
\bigskip

\noindent
{\it 3.3. The Fixation of Coordinates and of the Cut $\beta$ = 0}
\bigskip
Consider the trial configuration of figure 1.  We are free to
position the axis of coordinates in the simplest way; we shall make the
following choice.  The central line has two points $\mu = 0, \mu = \lambda_c$;
we use this as the $z$ axis. The bottom of the central line is the origin of
the coordinates.  Unless the configuration is axially symmetrical, we have
several
ways, for orienting the $x y$ coodinates. For instance if the $y z$ plane cuts
the $\lambda = 0$ line at P, the  orientation may be chosen so that $y_{_P}$
be
maximum. There may be several extrema of $y_{_P}$ and it does not matter which
one we take.
We shall only compare configurations with small differences.
The cut $\beta  = 0$  may be the ruled surface generated by lines parallel
to the $x$ axis and sliding on the central string.

In this way, we fix  uniquely  the relative positions of trial configurations
with respect to the coordinates and the
parametrization  of their points is:
 $$ 0 \le \alpha \le \alpha_M, \quad 0 \le \beta \le 2\pi,
    \quad 0\le \mu \le \lambda(\alpha) \eqno(3.24)$$
\bigskip
\noindent
{\it 3.4. Proof That the Labelling Is Mass and Circulation Preserving}
\bigskip
Consider a fluid element labeled $(\alpha, \beta, \mu) \equiv (\alpha^k)$ with
the coordinates $x,y,z$ or $\vec r$. Any displacement from  $\vec r$ to
$\vec r  + \Delta \vec r$ of that fluid element
($\Delta \alpha = \Delta \beta= \Delta \mu= 0$) is obviously mass preserving
since according to (3.5)
$$ \Delta(\rho d^{3}x) = \Delta d^{3}\alpha = 0    \eqno(3.25)$$
This shows incidentally that if we set
 $$ \Delta \vec r = \vec \xi(\vec r), \quad {\rm then} \quad
    \Delta \rho = - \rho \vec \nabla \cdot \vec \xi  \eqno(3.26)$$
Similarly from (3.4), the flux of the vorticity through any surface element
$d\vec S$ parametrized by $\alpha, \beta$ is
$$ \vec \omega \cdot d\vec S = d\alpha d\beta   \eqno(3.27)$$
which shows again that
$$ \Delta(\vec \omega \cdot d\vec S) = 0  \eqno(3.28)$$
or (see Arnold 1966, LBK1981)
$$ \Delta \vec \omega + {\rm rot}(\vec \omega \times \vec \xi)
    = (\vec \xi \cdot \vec \nabla)\vec \omega   \eqno(3.29)$$
\bigskip

\noindent
{\bf 4. Energy Principles}
\bigskip
We shall now show that the energy of steady flows is stationary compared
to the energy of any nearby trial configuration with the same total mass,
linear and angular
momentum and the same mass and vortex flux in displaced fluid elements.
\bigskip

\noindent
{\it 4.1. Some Differential Identities}
\bigskip

Let $F$ be a function of ($\alpha, \beta, \mu) \equiv \alpha^k$. Since
$$ {\partial \alpha^{l} \over \partial x^m}
   {\partial x^{m} \over \partial \alpha ^k}
   = \partial_{m}\alpha^{l}\partial_{\tilde k} x^m
   = \vec \nabla \alpha^{l} \cdot \partial_{\tilde k} \vec r
   = \delta^l_k,    \eqno(4.1)$$
it follows that
$$\Delta(\vec \nabla \alpha^{l})
   = - \vec \nabla \xi^{k} \partial _{k}\alpha ^l
   =  - \vec \nabla \vec \xi \cdot \vec \nabla \alpha ^l
 \eqno(4.2)$$
and therefore
$$ \Delta \vec \nabla F = - \vec \nabla \xi^{k} \partial _{k}F
   + \vec \nabla \Delta F = - \vec \nabla  \vec \xi \cdot \vec \nabla F +
   \vec \nabla \Delta F \eqno(4.3)$$
In particular, see (3.9),
$$ \eqalignno{
    \Delta \vec W = \alpha \Delta \vec \nabla \beta + \Delta \vec \nabla \nu
    &= - \vec \nabla \vec \xi \cdot \vec W
    + \vec \nabla \Delta \nu \cr
    &= \vec \nabla \vec W \cdot \vec \xi +
    \vec \nabla ( \Delta \nu - \vec W \cdot \vec \xi) &(4.4) \cr}$$
Where  $\Delta \nu$ is defined by varying (3.10)
$$ \Delta [\vec \nabla \cdot (\rho \vec \nabla \nu)]
    =  \Delta \vec \nabla \cdot [\rho ( - \alpha \vec \nabla\beta
    + \vec \eta_c)]  \eqno(4.5)$$
This defines $\Delta \nu$ by an elliptic equation of the same form as  $\nu$
itself
$$  \vec \nabla \cdot [\rho \vec \nabla (\Delta \nu)] = etc...     \eqno(4.6)$$
and therefore $\Delta \nu$ is defined with the same type of Green function as
$\nu$.
Another needed variation is that of $\Phi$ which, according to (2.1) and (3.5)
may be written
$$ \Phi = - G \int {d^{3} \alpha'
   \over |\vec r(\alpha) - \vec r^{~'}(\alpha')|}
    = - G \int {d^{3}\alpha' \over R}  \eqno(4.7)$$
Thus
$$ \Delta \Phi = - G \int \left(\vec \xi \cdot \vec \nabla {1 \over R}
    + \vec \xi \cdot \vec \nabla' {1 \over R}\right) d^{3}\alpha'  \eqno(4.8)$$
and therefore the variation of the gravitational potential energy
$$ \Delta \int { 1 \over 2} \Phi \rho d^{3} x = \Delta \int {1 \over 2} \Phi
   d^{3} \alpha = \int {1 \over 2} \Delta \Phi d^{3} \alpha
   = \int \vec \xi \cdot \vec \nabla \Phi \rho d^{3} r.  \eqno (4.9)$$
\vfill\eject
\noindent
{\it 4.2. The Variation of the Energy}
\bigskip
The energy of the flow
$$ E = \int \left[{1 \over 2} \vec W^{2} + \varepsilon(\rho) + {1 \over 2} \Phi
   \right]
   \rho d^{3}x  \eqno(4.10)$$
in which $\varepsilon (\rho)$ is the specific internal energy of the barotropic
fluid, related to the pressure and the specific enthalpy:
$$ \varepsilon(\rho) = h - {P \over \rho}  \qquad P = - \rho^{2}{\partial
   \varepsilon\over \partial \rho}   \eqno(4.11)$$
With $\Delta \vec W$ given in (4.4) and with
$\Delta \int {1 \over 2} \Phi \rho d^3 x$ in (4.9), one readily finds
the following identity for $\Delta E$,
$$ \eqalignno{
   \Delta E = \int \{ [(\vec W \cdot \vec \nabla) \vec W
   &+ \vec \nabla(h + \Phi)]\cdot \vec \xi \rho
   - \vec \nabla \cdot
   (\rho\vec W) (\Delta \nu - \vec W \cdot \vec \xi) \} d^{3}x \cr
   &+ \int \vec \nabla \cdot [(\Delta \nu - \vec W \cdot \vec \xi) \rho \vec W
   - P \vec \xi]d^3 x.   &(4.12) \cr} $$
\bigskip

\noindent
{\it 4.3. The Constrained Variational Identity and the Principle of Stationary
Energy}
\bigskip

We can now also calculate $\Delta \vec P$ and $\Delta \vec J$ which according
to (2.11) are given by
$$ \Delta \vec P = \int \Delta \vec W d^{3} \alpha \quad {\rm and}
   \quad \Delta \vec J = \int (\vec \xi \times \vec W
   + \vec r \times \Delta \vec W) d^{3} \alpha    \eqno(4.13)$$
Using (2.3) that defines $\vec U$, we then obtain a new identity:
$$ \eqalignno{
   \Delta E - \vec b \cdot \Delta \vec P - \vec  \Omega_c \cdot \Delta \vec J
    = &\int [\vec{\cal O} \cdot \vec \xi \rho +
    {\cal U} ( \vec W \cdot \vec \xi - \Delta \nu )
     ]d^{3} x  \cr &
    +\int_S [(\Delta \nu - \vec W \cdot \vec \xi) \rho \vec W
   - P \vec \xi] \cdot d\vec S,  &(4.14) \cr} $$
where $\vec{\cal O}$ and ${\cal U}$ have been  defined in (2.2) and (2.7),
respectively.

We see from (4.14) that if $\rho|_s=P|_s=0$, mass is preserved
(${\cal U}= 0$) and Euler's equations for steady flows in uniformly moving
coordinates
hold
($\vec {\cal O} = 0$) then the energy is stationary ($\Delta E = 0$) when
linear
and angular momenta are kept fixed ($\Delta \vec P = \Delta \vec J = 0$).
\bigskip

\noindent
{\it 4.4. Weak and Strong Principles of Minimum Energy}
\bigskip

Reciprocally if $ \Delta E = 0$ with $ \Delta \vec P = \Delta \vec J = 0$
(which
defines $\Delta \vec b$ and $\Delta \vec  \Omega_c$), then the following must
hold:

($\alpha$) Either $\nu$ is defined by mass conservation and the principle of
stationary energy provides Euler's equations $\vec {\cal O} = 0$,

($\beta$) Or $\nu$ is an additional variable and $\Delta E = 0$ if,
in addition to $\vec {\cal O} = 0$, we have mass conservation, ${\cal U} = 0$.

The principle of energy with four instead of three independent functions,
$\vec r$ and $\nu$, we shall call the weak energy principle, as opposed to that
with only three independent functions which we call the strong energy
principle.

As far as stationarity of energy is concerned, the distinction is of
no consequence. But it becomes important when we consider second order
variations and the principle of minimum energy. Second order variations of
(4.14) near the extremum $\Delta E = 0$ are given by
$$ \Delta^{2}E - \vec b \cdot \Delta^{2} \vec P - \vec  \Omega_c \cdot \Delta
^{2}
 \vec J = \int [\Delta \vec {\cal O} \cdot \vec \xi \rho -
 \Delta {\cal U} (\Delta \nu - \vec W \cdot \vec \xi)] d^{3}x
  \eqno(4.15)$$
Notice that with (2.9) and (2.10), the boundary term in (4.14) does not
contribute to $\Delta^2 E$. The relations
$\Delta^{2} \vec P = \Delta^{2} \vec J = 0$ define $\Delta^{2} \vec b$
and $\Delta^2 \vec  \Omega_c$. Stationary configurations are
stable if
$$ \Delta ^{2} E > 0. \eqno(4.16)$$
The strong energy principle reads then
$$ \Delta^{2} E_{strong}= \int \Delta \vec {\cal O} \cdot \vec \xi \rho d^{3}x
> 0
    \eqno(4.17)$$
which contains $\Delta \nu$ that must be obtained from (4.6).  The weak energy
principle is
$$\Delta ^{2} E_{weak} = \int [\Delta \vec {\cal O} \cdot \vec \xi \rho -
   \Delta {\cal U} (\Delta \nu - \vec W \cdot \vec \xi )] d^{3}x > 0
  \eqno(4.18)$$
and is manifestly simpler to apply since $\Delta \nu$ is here independent.
The weakness of this principle compared to (4.17) can be understood by the fact
that variations of the trial functions are mass preserving and isocirculational
but that the velocity trial field  $\vec W$ does not conserve mass.  Only
extremal  $\vec W$'s do.  Thus fluctuations in $\Delta^2 E$ would include
$\Delta \vec W$'s that do not necessarily satisfy the mass conservation
equation.  $\vec W + \Delta \vec W$ being less restricted than a real dynamical
$\vec W + \Delta \vec W$, some instabilities might show up that cannot exist
and would not show
up in a perturbation analysis.  For this reason this energy principle is
weaker.
But the weak principle may be helpful when the strong principle involves too
hard calculations.  Moreover, in numerical calculations, in which the computer
"searches" the minimum of $E$, one has to find the Green functions $G(\vec r,
\vec r \ ')$ for $\nu$ at every step
and that may be time consuming.  Therefore the energy principle with four
independent functions may indeed be useful.
\bigskip

\noindent
{\bf 5. Principles of Stationary Action}
\bigskip

\noindent
{\it 5.1. Introduction}
\bigskip
It may appear to make little sense to set up a Lagrangian formulation in terms
of variables
that take a priori account of all conservation laws of motion.  First the
dynamical equations insure automatically that the constants of the
motion are indeed constant.  Second, in hydrodynamics, fixation of the values
of constants of
motion does not reduce very much the number of independent variables but
complicate considerably the equations of motion.

However, it is important to show that our labelling defines proper Lagrange
variables, even if we are never going to use them, for the following reason.
An ordinary
Lagrangian  has the form $L = T - V$ while the energy $E = T + V$ ($T$ the
kinetic
energy, $V$ the potential energy).  If $T$ is purely quadratic in the time
derivatives of the Lagrange variables,  $\Delta^2 E > 0 $ is a necessary and
sufficient condition of stability.  It is however well known that Lagrangians
 subject to non-holonomic constraints, like circulation conservation or fixed
linear momentum, contain additional terms, so called gyroscopic terms G
, that are {\it linear} in time derivatives. The form of $L$ is then $T + G -
V$, the energy, however, is still $T + V $. In those
circumstances, $ \Delta^2 E > 0 $ is only a sufficient condition of stability,
not a necessary one. In our representation in which
not only fixed $\vec J$ and $ \vec P $ but also fixed circulation and mass
conservation have been incorporated, we may have lots of
gyroscopic terms in the Lagrangian. It is therefore useful to consult the form
of $L$ and to see if there are gyroscopic terms so as to
know when our energy principles give necessary and sufficient conditions of
stability or only sufficient ones.
For this reason we have first to proof that the Lagrangians of our flows
 provide indeed Euler's equations.
\bigskip

\noindent
{\it 5.2. Time Dependent Lagrange Variables and Dynamical Equations}
\bigskip
We use the same Lagrange variables $\alpha^k$ introduced in section 3. Now,
however, they depend also on the time:  $ \alpha^k = \alpha^k(\vec r, t) $.
Surfaces of constant $ \alpha^k$ define a velocity $ \vec w$ which
satisfy the equations:
$$ \alpha^k(\vec r + \vec w dt,t + dt) = \alpha^k(\vec r,t); \eqno(5.1)$$
in the limit $dt \to 0$:
$$ \dot \alpha^k + \vec w \cdot \vec \nabla \alpha^k = 0 .\eqno(5.2)$$
Since  {\it arbitrary} displacements of a fluid element with labels
$\alpha^k$ (see section 3.4) are both mass preserving and isocirculational it
follows from (5.2), (3.4) and (3.6) that mass and vortex strength are both
preserved along a motion with velocity $ \vec w $:
$$ \dot \rho + \vec \nabla \cdot ( \rho \vec w) = 0 \eqno(5.3)$$
and
$$ \dot {\vec \omega} +  rot (\vec \omega \times \vec w) = 0 \eqno(5.4)$$
There exists a useful explicit expression for $\vec w$ that simplifies
various formulas; consider the following identity obtained from $ \vec r (t,
\alpha^k)$:
$$ \vec r \equiv \vec r [t, \alpha^k( \vec r,t)] \eqno(5.5) $$
Since the right hand side must be independent of $t$, we have:
$$ ({ \partial \vec r \over \partial t})_{r} = ({ \partial \vec r \over
\partial t})_{\alpha} + \dot \alpha^k { \partial \vec r \over \partial
\alpha^k}
\eqno(5.6)$$
extracting $ \dot \alpha^k $ from (5.2) and replacing it in
(5.6) gives then:
$$ ({ \partial \vec r \over \partial t})_{\alpha} = \vec w \eqno(5.7)$$
and, quite generally,
$$ ({ \partial F(t, \vec r) \over \partial t})_{\alpha} = \dot F + \vec w \cdot
\vec \nabla F \eqno(5.8)$$
The "absolute" velocity of the fluid is, say, $ \vec W $ plus the velocity of
the vortex lines $ \vec w$:
$$ \vec v = \vec w + \vec W \eqno(5.9)$$
The relative velocity is similarly:
$$ \vec u = \vec w + \vec U \eqno(5.10)$$
$ \vec W $ and $\vec U$  are related by eq. (2.3). The equation of mass
conservation along the motion of the fluid must be
$$ \dot \rho + \vec \nabla \cdot ( \rho \vec u) = 0 \eqno(5.11)$$
and (5.3), with (5.10) and (5.11), imply that $ \vec U $ satisfies equation
(2.7) again:
$$ {\cal U} \equiv \vec \nabla\cdot (\rho \vec U) = 0.      \eqno(5.12)$$
The dynamical equations of motion are slightly different from (2.2) and are
well
known:
$$ \vec {\cal O}_{{ D}} \equiv \dot {\vec v} + (\vec u \cdot \vec \nabla)\vec v
+
   \vec  \Omega_c \times \vec v + \vec \nabla (h + \Phi) = 0,  \eqno(5.13)$$
This familiar equation may also be written in a slightly less familiar form
which will be useful; with (5.8) and (5.10), we have
$$ \vec {\cal O}_{    D} \equiv ({ \partial \vec v \over \partial t})_{\alpha}
+ (\vec U \cdot \vec \nabla)\vec v +  \vec  \Omega_c \times \vec v + \vec
\nabla (h + \Phi) = 0,  \eqno(5.14)$$
In stationary flows $ \vec w =0 $ ,$ \vec v = \vec W $ and $({ \partial \vec W
\over \partial t})_{\alpha} = 0$ ; $ \vec {\cal O}_{    D}$ becomes then the
$\vec {\cal O}$ of (2.2).
\bigskip

\noindent
{\it 5.3. The Action}
\bigskip
The Action of the system is given (see KLB 85) by:
$$ \eqalignno{ A = \int ^{t_1}_{t_0} L dt \equiv \int \int [ {1 \over 2} \vec
w^{2} - ({1 \over 2}
   \vec W^{2} + \varepsilon + {1 \over 2} \Phi)
   ] \rho d^{3}x dt = \int \int [ {1 \over 2} \vec w^{2} \rho d^{3}x - E(t)]dt
\ \
&(5.15) \cr} $$
Notice the minus sign and the fact that $E$ appears here effectively as the
potential. The Lagrangian of isocirculational flows is actually a
Routhian that incorporates mass and circulation conservation. With (5.9) we may
write $L$ as this:
$$ L =
 \int [\vec w \cdot \vec v - \left({1 \over 2}
   \vec v^{2} + \varepsilon + {1 \over 2} \Phi\right)]\rho d^{3}x \eqno(5.16)$$
which is of great help in calculations. The Lagrange variables are thus
$(\alpha, \beta, \mu) = \alpha^k$ as in stationary flows; $ \vec v$, instead of
$\vec W$,
has now a Clebsch form:
$$ \vec v = \alpha \vec \nabla \beta + \vec \nabla \nu. \eqno(5.17)$$
where $\nu$ is a single valued functional of $\alpha^k$ defined by (5.12) and
is  linear in $(\dot \alpha^k, \vec b,
\vec \Omega_c)$  because $ \vec w $ is linear and homogeneuos in
$ \dot \alpha^k $.  Quantities like $\varepsilon$ and $ \Phi $ depend on $ \rho
$ only which is a functional of $ \alpha^k$. The Action is thus a functional of
$ \alpha^k $ , quadratic in $ \dot \alpha^k $ but not quadratic homogeneous.
Notice that $L$ is apparently linear only in $ \dot \alpha^k$ through $\vec w$.
However, $L$ is {\it quadratic} in $\nu$ which is also {\it linear} in $ \vec
w$, $\vec b$ and $\vec \Omega_c$.
The constants $\vec b$ and $\vec \Omega_c$ are defined by the values of linear
and angular momentum which are now given by:
$$ \vec P  = \int \vec v \rho d^3 x = \int (\alpha \vec \nabla \beta + \vec
\nabla \nu) \rho d^3 x= 0, \eqno(5.18a)$$
and
$$  \vec J = \int \vec r \times \vec v \rho d^3x = \int \vec r \times (\alpha
\vec \nabla \beta + \vec \nabla \nu) \rho d^3x = \vec J_0;  \eqno(5.18b)$$
$ \vec b$ and $ \vec \Omega_c$ are thus equally linear in $ \dot \alpha^k$
through $\nu$.
\bigskip

\noindent
{\it 5.4. Variational Identities for Calculating $\Delta A$}
\bigskip
Let us now make small displacements $\vec \xi $ of fluid elements and calculate
the changes $\Delta A$, $\Delta \vec P$ and $\Delta \vec J$ in $A, \vec P$ and
$ \vec J $ . Much of the calculation has actually been done in section~4.
Indeed notice first that $  E(t), \vec P$ and $ \vec J$ are similiar to $E$ in
(4.10), $\vec P $ in
(2.11b) and $ \vec J$ in (2.11c) with $ \vec v $ replacing $ \vec W$.
Since $ \vec v $, in time dependent flows, and $ \vec W $, in steady flows, are
  both represented by $\alpha \vec \nabla \beta + \vec \nabla \nu$ , we obtain,
straightaway $ \Delta  E(t) - \vec b \cdot \Delta \vec P - \vec \Omega_c \cdot
\Delta \vec J $
$ \vec U $ by $\vec u $:
$$ \eqalignno{ & \Delta  E(t) - \vec b \cdot \Delta \vec P - \vec  \Omega_c
\cdot \Delta \vec J \cr & =  \int \{ [ (\vec u \cdot \vec \nabla)\vec v +
   \vec  \Omega_c \times \vec v + \vec \nabla (h + \Phi)]
    \cdot \vec \xi \rho +
    \vec \nabla \cdot ( \rho \vec u) ( \vec v \cdot \vec \xi - \Delta \nu )
     \} d^{3} x  \cr & \ \ \ \ \
    +\int_S [(\Delta \nu - \vec v \cdot \vec \xi) \rho \vec u
   - P \vec \xi] \cdot d\vec S,  &(5.19) \cr} $$
Notice that (5.19) is a variational identity; $\nu$ has been treated as an
independant function. To calculate $ \Delta A $ there remains, however, some
work to do on the $ \vec v
\cdot \vec w $ part of $L$. Following (5.7) and (5.17)
$$ \Delta (\vec w \cdot \vec v) = \Delta ({\partial \vec r \over \partial t})_{
\alpha} \cdot \vec v + \vec w \cdot \Delta (\alpha \vec \nabla \beta + \vec
\nabla \nu) \eqno(5.20) $$
We can use (4.4) to rewrite $\Delta (\alpha \vec \nabla \beta + \vec \nabla
\nu) $and obtain
$$ \Delta (\vec w \cdot \vec v) = ({\partial \vec \xi \over \partial
t})_{\alpha} \cdot \vec v + (\vec w \cdot \vec \nabla) \vec v \cdot \vec \xi +
\vec w
\cdot \vec \nabla (\Delta \nu - \vec v \cdot \vec \xi) \eqno(5.21) $$
so that, with some integration by part and, with the help of (5.8), we can
write
$$ \Delta \int \vec w \cdot \vec v \rho d^3 x =  {\partial (\int \vec v \cdot
\vec \xi \rho d^3 x) \over \partial t} + \int [
    -\dot {\vec v} \cdot \vec \xi \rho +
    \vec \nabla \cdot ( \rho \vec w)( \vec v \cdot \vec \xi - \Delta \nu )
      ]d^{3} x
    +\int_S (\Delta \nu - \vec v \cdot \vec \xi) \rho \vec w \cdot d\vec S,
 \eqno (5.22) $$
Thus, $ \int [\Delta L + \vec b \cdot \Delta \vec P + \vec  \Omega_c \cdot
\Delta \vec J] dt$ is given by (5.22) minus (5.19); in writing the result we
shall use
$ \vec {\cal O}_{    D} $ defined in (5.13) and $ {\cal U} $ in (5.12):
$$ \eqalignno{ &\Delta A + \int(\vec b \cdot \Delta \vec P + \vec  \Omega_c
\cdot \Delta \vec J) dt = (\int \vec v \cdot \vec \xi \rho d^3 x)|_{t_0}^{t_1}
    - \int \int [ \vec {\cal O}_{ D}
    \cdot \vec \xi \rho +
    {\cal U}( \vec v \cdot \vec \xi - \Delta \nu )
      ]d^{3} x dt  \cr &
 - \int \int_S [(\Delta \nu -  \vec v \cdot \vec \xi) \rho \vec W
   - P \vec \xi] \cdot d\vec S dt ,  &(5.23) \cr} $$
\bigskip

\noindent
{\it 5.5 Weak and Strong Principles of Stationary Action}
\bigskip

Identity (5.23) leads straightaway to the following results. If $ \rho|_s =
P|_s=0$, mass is preserved (${\cal U} = 0$) and Euler's equations hold ($ \vec
{\cal O}_{D} = 0$) then the Action is  stationary  ($ \Delta A = 0 $) when
linear and angular moment
Reciprocally, if $ \Delta A = 0 $ with $ \Delta \vec P = \Delta \vec J = 0 $
(which define $ \Delta \vec b $ and $ \Delta \vec \Omega_c$ ) then the
following must hold:

$ (\alpha)$ Either $\nu$ is defined by  ${\cal U} = 0$ and the principle of
stationary Action provides Euler's equation $ \vec {\cal O}_{ D} = 0$ .

$ (\beta)$ Or $\nu$ is an additional variable and A is stationary if in
addition to
$ \vec {\cal O}_{D} = 0$, we have mass conservation ${\cal U} = 0$.
The principle of stationary Action, with four independent functions, $
\alpha^k$ and $\nu$,
instead of three, we call the weak principle of stationary Action. It is the
dynamical counterpart to the weak energy principle.
\bigskip

\noindent
{\bf 6. Gyroscopic Terms for Small Deviations from Steady Flows}
\bigskip
\noindent
{\it 6.1 The Action for Small Deviations}
\bigskip
Small time dependent deviations from a stable flow are given by Euler's
linearized
equations whose Action is the second order variation $\Delta^2 A$ of $A$
calculated at the stationary "point". Let $ \vec r_0 (t,\alpha^k) $ be the
coordinates of the fluid
element of a particular stationary flow and $ \vec r_0 + \vec \xi (t,\alpha^k)
$ the coordinates in the perturbed flow. The gyroscopic terms in $\Delta^2 A$
are the bilinear  functionals of $ \vec \xi $ and $ \dot {\vec \xi} $.
The gyroscopic terms of the fully constrained Action are given by the $ \vec
\xi $ , $ \dot {\vec \xi} $ terms in $\int [\Delta^2 L + \vec b \cdot \Delta^2
\vec P + \vec  \Omega_c \cdot \Delta^2 \vec J] dt $ as evaluated at the
stationary point. Let  the
$$ \vec w_0 =0, \qquad \vec v_0 = \vec W_0, \qquad \vec u_0 = \vec U_0, \qquad
\vec \nabla \cdot (
\rho_0 \vec U_0) =0  \qquad etc... \eqno(6.1)$$
Notice that:
$$ ({\partial \vec \xi \over \partial t})_{\alpha}|_0 = \dot {\vec \xi}
 + (\vec w_0 \cdot \vec \nabla) \vec \xi = \dot {\vec \xi} \eqno(6.2)$$
With (6.1) and (6.2) we readily obtain $ \Delta^2 L + etc...$ from (5.23) at
the
stationary "point". Notice that because of (2.9) and (2.10), the second
variations have no surface terms.  Since the terms at $t=t_0$ or $t_1$ do not
contribute either, we have:
$$ [\Delta^2 A + \int ^{t_1}_{t_0} (\vec b \cdot \Delta^2 \vec P + \vec
\Omega_c \cdot \Delta^2 \vec J) dt]_0 = \int \int [- \Delta \vec {\cal O}_{
D} \cdot \vec \xi \rho + \Delta {\cal U}( \Delta \nu - \vec v \cdot \vec \xi
)]_0  dt d^3 x
\eqno(6.3) $$
in which
$$ \Delta  {\cal U}|_0 = \Delta [\vec \nabla \cdot (\rho \vec U)]|_0 = \vec
\nabla \cdot [ \rho_0 ( \vec \nabla \Delta \nu - \dot {\vec \xi} - (\vec \nabla
\vec\xi \cdot \vec U_0 + \vec U_0 \cdot \vec \nabla \vec \xi) - \vec \nabla
\vec \xi\cdot \vec \eta$$
where $ \vec \eta_c $ is the vector defined in (2.3); of $\Delta \vec {\cal
O}_{    D}$ we write only those parts susceptible to contribute to the
Gyroscopic term:
$$  \eqalignno{& ( \int \int - \Delta \vec {\cal O}_{D} \cdot \vec \xi \rho_0
dt d^3 x)_{Gyro} = \int \int \{ \dot {\vec \xi} \cdot (\vec \nabla \Delta \nu -
\vec  \nabla \vec \xi \cdot \vec W_0) + [(\dot {\vec \xi} - \vec \nabla \Delta
\nu+ \Delta \vec{ } &XXX\cr}$$

\bigskip
\noindent
{\it 6.2 Gyroscopic Term for the Strong Principle.}
\bigskip
In the strong energy and Action principle, $\nu$ is defined by $ {\cal U} = 0$
and $ \Delta \nu $ is defined by $ \Delta {\cal U}|_0 = 0$. $ \Delta \nu $ may
be decomposed into a "dynamical contribution" $ \Delta \nu_D$ which is zero
when $ \dot {\vec \xi} = 0 $ and a "steady" part $ \Delta \nu_S$:

$$ \Delta \nu =  \Delta \nu_D +  \Delta \nu_S \eqno(6.6)$$

Similarly, $ \vec \eta_c $ has a dynamical and steady contribution (see our
remarks at the end of section 5.3) and we write

$$ \Delta \vec \eta_c = \Delta \vec \eta_{cD} + \Delta \vec \eta_{cS}
\eqno(6.7)$$
$ \Delta {\cal U}|_0 = 0$ gives the following two equations for $ \Delta \nu_D$
and $\Delta \nu_S$ deduced from (6.4)
$$  \vec \nabla \cdot ( \rho_0  \vec \nabla \Delta \nu_D ) =  \vec \nabla \cdot
[\rho_0 (\dot {\vec \xi} + \Delta \vec \eta_{cD})] \eqno(6.8a)$$
$$  \vec \nabla \cdot ( \rho_0  \vec \nabla \Delta \nu_S ) =  \vec \nabla \cdot
[\rho_0 ( \vec U_0 \cdot \vec \nabla \vec \xi + \vec \nabla \vec \xi \cdot \vec
U_0) + \rho_0 \vec \nabla \vec \xi \cdot \vec \eta_c + \Delta \vec \eta_{cS})]
\eqno(6.8b)$$

Gyroscopic terms in (6.3) will only come from $ \Delta \vec {\cal O}_{    D} $
since
$ \Delta {\cal U}|_0 = 0$. Using eq. (6.8) the bilinear terms of $ [\Delta^2 A
+ \int ^{t_1}_{t_0}( \vec b \cdot \Delta^2 \vec P + \vec  \Omega_c \cdot
\Delta^2 \vec J ) dt]_0 $, which we denote by $ \Delta^2 G_{strong} $ are

$$  \eqalignno{ &\Delta^2 G_{strong} = 2 \int \int [\vec \nabla
 \Delta \nu_D \cdot \vec \nabla  \Delta \nu_S  - \dot {\vec \xi} \cdot \vec
\nabla \vec \xi \cdot \vec W_0] \rho_0
 d^3 x dt \cr &- \int \int [\Delta \vec \eta_{cD} \cdot \vec \nabla  \Delta
\nu_S + \Delta \vec \eta_{cS} \cdot \vec \nabla  \Delta \nu_D
+ \Delta \vec \eta_{cD} \cdot \vec \nabla \vec \xi \cdot \vec W_0
\cr &+ (\Delta \vec \Omega_{cD} \times \vec W_0 + \vec \Omega_c \times \vec
\nabla  \Delta \nu_D) \cdot \vec \xi]  \rho_0 d^3 x dt,
&(6.9) \cr} $$

This expression becomes somewhat simpler for motions  that are steady in
inertial coordinates ($ \vec \eta_c = 0 $), when, for some reason, $\Delta \vec
\eta_c $ does not contribute either:
$$  \Delta^2 G_{strong} = 2 \int \int  [\vec \nabla
 \Delta \nu_D \cdot \vec \nabla  \Delta \nu_S - \dot {\vec \xi} \cdot \vec
\nabla \vec \xi \cdot \vec W_0 ]  \rho_0 d^3 x dt \eqno(6.10)$$
We conclude that the strong energy principle provides a {\it necessary and
sufficient}
condition of stability if:
$$ \Delta^2 E_{strong} > 0 \qquad and \qquad \Delta^2 G_{strong} = 0
\eqno(6.11)$$
\vfill\eject

\noindent
{\it 6.3 Gyroscopic Term and the Weak Energy Principle.}
\bigskip
The weak Lagrangian is linear in time derivatives; $L$ is of the form  $G-V$
since $T=0$. The perturbed Action has no quadratic term in $ \dot {\vec \xi}$
which  appears only in $\Delta^2 G_{weak}$. As a result, if $\Delta^2 G_{weak}
\ne 0$,
 $\Delta^2 E_{weak} > 0$ is certainly a sufficient condition of stability (this
can easily be shown). However, if $\Delta^2 G_{weak} = 0$ and $\Delta^2
E_{weak} > 0$, the linearized equation of motion have no solution at all! Usual
arguments about stabili

 It is
 than still interesting to obtain $\Delta^2 G_{weak}$ for, however, very
different reasons than we wanted $\Delta^2 G_{strong}$. In the weak Action
principle,
$ \vec P$ and $ \vec J $ do not define $ \vec b$ and $\vec \Omega_c$ [see
(5.18)]
because $\nu$ is independent of them.
 Instead of describing the motion in moving coordinates, we rather stay in
inertial coordinates and fix $ \vec P = \vec J - \vec J_0 = 0 $ with Lagrange
multipliers, say, $ \vec b(t) $ and $ \vec \Omega_c (t) $. The Action is then
$$ A ^{\dagger} = A + \int [ \vec b(t) \cdot \vec P + \vec \Omega_c(t) \cdot
(\vec J - \vec J_0) ] dt \eqno(6.12)$$
Among all the dynamical perturbations, we consider only those for which $ \vec
P = \vec J - \vec J_0 = 0 $. The formal expression for $ \Delta^2 A^{\dagger}$
is than exactly the same as the right hand side of (6.3). The gyroscopic term
$\Delta^2 G_{weak}$
$$ (\Delta^2 G )_{weak} = \int \int [(2 \dot {\vec \xi} + \Delta \vec
\eta_{cD})
\cdot (\vec \nabla \Delta \nu - \vec \nabla \vec \xi \cdot \vec W_0) - \Delta
\vec \Omega_{cD} \times \vec W_0 \cdot \vec \xi] \rho_0 d^3 x dt \eqno(6.13) $$
We conclude that a sufficient condition for stability is
$$ \Delta^2 E_{weak} > 0 \qquad and \qquad \Delta^2 G_{weak} \ne 0 \eqno(6.14)
$$
\bigskip
This work was started when S.I. visited the Racah Institute of Physics,
Hebrew University of Jerusalem in 1987.  He acknowledges its support and
hospitality.
\vfill\eject

\noindent
{\bf References}

\refs

Antonov, V.A. 1962, {\it Vestnik Leningrad Univ.}, {\bf 7}, 135 translated
in Goodman, J. and Hut, P. 1985, in
{\it Dynamics of Star Clusters}, IAU Symposium
No. 113 (Reidel)

Arnold, V.I. 1966, {\it J. M\'ec.}, {\bf 5}, 29.

Binney, J. \& Tremaine, S. 1987, {\it Galactic Dynamics}, Chap. 5 (Princeton
University Press).

Chandrasekhar, S. 1969, {\it Ellipsoidal Figures of Equilibrium} (Yale
University Press).

Katz, J. \& Lynden-Bell, D. 1985, {\it Geophys. Astrophys. Fluid Dynamics},
{\bf 33}, 1.

Lebovitz, N.R. 1965, {\it Astrophys. J.}, {\bf 142}, 229.

Lin, C.C. 1963, {\it \lq Liquid Helium'} in {\it Proc. Int. School Phys. XXI}
(Academic Press).

Lynden-Bell, D. \& Katz, J. 1981, {\it Proc. Roy. Soc. London}, A{\bf 378},
179.

Seliger, R.L. \& Whitham, G.B. 1968, {\it Proc. Roy. Soc. London},
A{\bf 305}, 1.

Serrin, J. 1959, {\it \lq Mathematical Principles of Classical Fluid
Mechanics'} in {\it Handbuch der Physik}, {\bf 8}, 148.

Simo J.C., Lewis D. \& Marsden J.E. 1991, {\it Arch. Rational Mech. Anal.} {\bf
115}, 15.

\endrefs
\vfill\eject

\centerline{FIGURE CAPTIONS}

\refs

Fig. 1.  Representation of a trial configuration indicating a surface of
constant load $\lambda=\lambda_0$, a cut of constant $\beta=\beta_0$ hanging
on the central ``string''
and the line $\lambda=0$.  The point $P$, at a shortest
distance from the $z$-axis helps to define the orientation of the $x,~y$
plane.

\endrefs

\bye